\documentclass[12pt]{iopart}
\usepackage{dcolumn}
\usepackage{graphicx}
\usepackage{float}
\usepackage{bm}
\usepackage[usenames]{color}

\newcommand{\ket}[1]{|#1\rangle}

\usepackage{iopams}
\begin{document}
\title{Entanglement Dynamics of Two Independent Cavity-Embedded Quantum Dots}

\author{B. Bellomo$^1$, G. Compagno$^1$, R. Lo Franco$^1$, A. Ridolfo$^2$, and S. Savasta$^2$}
\address{$^1$CNISM \& Dipartimento di Scienze Fisiche ed Astronomiche, Universit\`a di Palermo, via Archirafi 36, 90123 Palermo, Italy \\
$^2$Dipartimento di Fisica della Materia e Ingegneria Elettronica, Universit\`{a} di Messina Salita Sperone 31, I-98166 Messina, Italy} 
\ead{bruno.bellomo@fisica.unipa.it}
\date{\today}

\begin{abstract}

{We investigate the dynamical behavior of entanglement in a system made by two solid-state emitters, as two quantum dots, embedded in two separated micro-cavities.
In these solid-state systems, in addition to the coupling with the cavity mode, the emitter is coupled to a continuum of leaky modes providing additional losses and it is also subject to a phonon-induced pure dephasing mechanism.
We model this physical configuration as a multipartite system composed by two independent parts each containing a qubit embedded in a single-mode cavity, exposed to cavity losses, spontaneous emission and pure dephasing.  We study the time evolution of entanglement of this multipartite open system finally applying this theoretical framework to the case of  currently available solid-state quantum dots in micro-cavities.}

\end{abstract}


\maketitle

\section{Introduction}

Cavity quantum electrodynamics (CQED) deals with the interaction among photons confined in a reflective micro-cavity and atoms or other particles. When a two-level system (for example, a two-level atom) is strongly coupled to a cavity mode \cite{Scully}, it is possible to realize important quantum information processing tasks, such as controlled coherent coupling and entanglement of distinguishable quantum systems \cite{Ima,Monroe,Michler}. In this respect solid-state devices, and in particular semiconductor quantum dots (QDs), utilized as "artificial atoms", are one of the most promising architectures for the possibility of miniaturization, electrical injection, control and scalability.
Recently, thanks to impressive progress in the technology of solid-state microcavities,  substantial advances have been made towards these goals. The strong coupling regime has been reached for the excitonic transition of quantum dots (QDs) \cite{Reith,Yoshie,Peter,Press,Hennessy}, and nanocrystals  \cite{nanocr1}coupled to optical semiconductor cavities, as well as for superconducting qubits coupled to microwave cavities \cite{Wallraff}. In all of these systems, the cavity-mode quality  factor (Q) can be very large while solid-state emitters are intrinsically coupled to the matrix they are embedded in. In fact, decoherence and phase relaxation unavoidably broaden any transition between the discrete states of such artificial atoms.
High experimental performances are required to realize quantum processors and it is thus important to establish how long a sufficient degree of entanglement can be maintained in spite of losses, decoherence and noise. Implications are the possibility to store entangled states in solid-state memories and entanglement preservation during local operations in quantum algorithms~\cite{nielsenchuang}.
Understanding how the entanglement is transferred from, e.g., a pair of independent initially entangled qubits to reservoirs has motivated several contributions in recent years \cite{lopez2008PRL,bai2009PRA}.
The aspect that has mostly drawn attention is the possibility of a complete disappearance of the entanglement between the qubits at finite times  \cite{yu2009Science}.  The occurrence of this phenomenon, termed ''entanglement sudden death'' (ESD) \cite{yu2004PRL} and of entanglement revivals  \cite{bellomo2007PRL} have been shown in  quantum optics experiments \cite{almeida2007Science,Jin-ShiXuPRL}. Entanglement transfer from atoms to cavity modes leading to entanglement revivals has also been studied \cite{Lopez2010PRA}. The effects on the dynamics of entanglement of cavity losses, spontaneous emission and pure dephasing for two-qubit systems have been extensively investigated \cite{Huang2007PRA,bellomo2008PRA}. The general problem of the dynamics of entanglement in the simultaneous presence of more than one noise has been also studied finding that, for composite systems, the additivity of decay rates of single noises is not maintained \cite{yu2006PRL}. The main aim of this paper is to investigate the dynamical behavior of entanglement for a realistic implementation for quantum computation of two qubits in separated cavities where several sources of noise are present.

Here we  consider the entanglement dynamics in a system composed by two initially correlated solid-state emitters each strongly coupled to a lossy cavity interacting with its reservoir. We include the unavoidable losses due to spontaneous emission into external electromagnetic modes distinct from the lossy single-mode of the cavity. These, always present, additional losses arise from the coupling of the emitter to a continuum of leaky modes  \cite{Auffeves}. We also include pure dephasing noise which plays a relevant role in solid state quantum emitters. For example, a QD interacts with the phonons of the matrix it is embedded in, giving rise to sidebands in addition to the so-called zero-phonon line (ZPL). At sufficiently low temperature yet, the emission in the ZPL remains predominant, allowing to model these systems as effective two-level systems subject to additional pure dephasing \cite{Auffeves}. We also study the entanglement transfer from the two-emitter system to the cavity modes. The input-output relations for optical cavities \cite{wallsbook} show that the entanglement between cavity modes can in principle be measured experimentally  by collecting photons escaping the cavities. Such two-cavity entanglement dynamics could thus be exploited to monitor the entanglement dynamics of cavity-embedded solid state emitters.

The paper is structured as follows. In Sec.~\ref{par:Model} we present and solve a model for the physical configuration described above. In Sec.~\ref{sec:entanglement-evolution} we explore the dynamical behavior of entanglement for quite general values of the physical parameters of the system. In Sec.~\ref{par:Application} we specialize to the case of currently available solid-state quantum dots in micro-cavities giving the characteristic lifetimes of entanglement. In Sec.~\ref{par:Conclusions} we summarize our results.

\section{Model}\label{par:Model}

Our system is composed by two noninteracting subsystems ($S=A,B$), each consisting in a qubit (two-level emitter) $q_S$ coupled to a single-mode cavity $c_S$ in turn interacting with an external reservoir $r_S$ (see Fig.~1). The Hamiltonian of the total system is thus given by the sum of the Hamiltonians of the two noninteracting subsystems
 \begin{eqnarray}\label{H totale}
     H_{\rm tot} &=& H_{\rm A} + H_{\rm B}.
\end{eqnarray}
\begin{figure}
\begin{center}
\includegraphics[width=7.4 cm, height=4.5 cm]{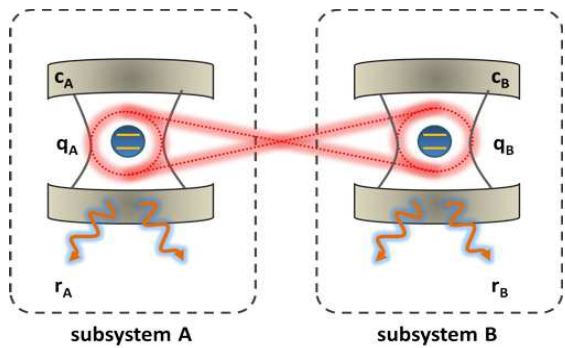}
\end{center}
\caption{\label{fig:system}\footnotesize (Color online) Schematic representation of the four-partite system ($q_A$, $c_A$; $q_B$, $c_B$)}. The two qubits $q_A$ and $q_B$ are initially entangled.
\end{figure}
In each subsystem, we distinguish the bipartite system made by the qubit and the cavity from the reservoir $r$. The Hamiltonian $H_S$ of each part $S=A,B$ reads like (we omit index $S$)
 \begin{equation}\label{H singola parte}
    H= H_{\rm o}+H_{\rm r}+H_{\rm i} ,\qquad
     H_{\rm o} =    \frac{1}{2} \omega_{\rm 0} \sigma_{\rm z}+ \omega_{\rm c}\, a^\dag a + g ( a^\dag\, \sigma_-  +  a\, \sigma_+)\,,
\end{equation}
where $H_{\rm o}$ describes, in the rotating wave approximation (RWA), the qubit-cavity system, $g$ is the coupling constant between qubit and cavity, $\sigma_{\rm z}$ denotes the usual diagonal Pauli matrix, $\sigma_{\pm}$ are the two-level raising and lowering operators, $a$ and $a^\dag$  the annihilation and creation operators for the cavity mode. We observe that this Hamiltonian model is valid for values of $g$ smaller than $\omega_0 \sim \omega_c$ \cite{werlang2008PRA}.
$H_{\rm r}$ describes the external environment responsible of the different noise sources which affect the qubit-cavity system and $H_{\rm i}$ the interaction of the latter with the environment. The realistic conditions present in a system composed by quantum dots embedded in microcavities are modeled as three noise sources: cavity losses, qubit spontaneous emission and pure dephasing mechanisms.

In the usual master equation approach, considering the Markov approximation and an infinite number of bath oscillators, we can describe the dynamics of the qubit-cavity system by
\begin{equation} \label{master equation}
	\frac{d}{dt} \rho = i[\rho, H_{\rm o}] + {\cal L}\, \rho\, .
\end{equation}
The Markovian processes are described by the Liouvillian term, that consist of 3 parts:
\begin{equation}
	{\cal L} = {\cal L}_{cav} + {\cal L}_{SE}+ {\cal L}_D\, .
\end{equation}
${\cal L}_{cav}$ describes the cavity losses of photon in the reservoir modes and is expressed by the following form  \cite{nielsenchuang}:
\begin{equation}
	{\cal L}_{cav} =
	 \frac{\gamma_{\rm c}}{2} (2 a \rho a^\dag - a^\dag a \rho - \rho a^\dag a)\, ,
\end{equation}
being $\gamma_{\rm c}$ the photon escape rate from the cavity to free space.
In addition, the qubit is subjected to decay \emph{via} spontaneous emission and losses of coherence.
${\cal L}_{SE}$ describes spontaneous emission in the leaky modes: all the available light modes except the cavity one  \cite{nielsenchuang},
\begin{equation}
	{\cal L}_{SE} = \frac{\gamma_{\rm q}}{2}(2 \sigma_- \rho \sigma_+ - \sigma_+\sigma_- \rho - \rho  \sigma_+\sigma_-)\, .
\end{equation}
${\cal L}_{cav}$ and ${\cal L}_{SE}$ are based on Hamiltonians in RWA and thus are not valid for arbitrarily large values of the decay rates $\gamma_{\rm c}$ and $\gamma_{\rm q}$ \cite{werlang2008PRA}. Finally,
\begin{equation}
{\cal L}_D = \frac{\gamma_{\rm d}}{4} (\sigma_z \rho \sigma_z - \rho)
\end{equation}
describes pure dephasing processes \cite{nielsenchuang}.
Although the three noises are treated in the Markovian limit, the reduced dynamics of the qubit-cavity system can present non-Markovian behavior depending on the strength of the coupling constant $g$ with respect to the various decay rates. We shall comment quantitatively on this point after Eq.~(\ref{p and q}).

The optical cavity is an open quantum system, cavity photons can escape it and propagate into free space on along an optical fibre until they eventually reach another distant quantum system or can be detected. The quasimode approach is able to describe in a direct way  such a behavior. A relationship between the external fields and the intracavity field may be obtained \cite{wallsbook,gardiner1985PRA} in the limit of continuous spectrum. If the coupling constant $\kappa(\omega)$ between cavity and external bosonic modes is independent of frequency over a band of frequencies about the characteristic frequency $\omega_c$, $\kappa(\omega)\approx \kappa$:
\begin{equation}
	a_{\rm out}(t) + a_{\rm in}(t) = \sqrt{\gamma_{\rm c}} a(t)\, ,
\label{eq:}
\end{equation}
where the operators $a_{\rm out}$ and $a_{\rm in}$ are related to the reservoir operators and $\gamma_{\rm c} = 2\pi \kappa^2$.
Throughout this paper we will consider the case of no input photons, hence, once known the quantum state for the cavity mode, it will be possible to calculate expectation values and correlation functions for output photons that can be measured experimentally or be used as input to transmit the entanglement to distant quantum systems.

\subsection{Procedure}

Being the two subsystems noninteracting, they evolve independently so that we can analyze the dynamics of only one subsystem and in turn obtaining the evolution of the total four-partite open system \cite{bellomo2007PRL}.  We will consider initial states with zero cavity photons and at most one excitation in each qubit.
Eq.~(\ref{master equation}) allows us to compute the joint evolution of the total four-partite open system starting from an arbitrary initial state where only one excitation is present in each atom. From the knowledge of the evolved density matrix, it will be possible to investigate the various reduced dynamics of the total system, for example that of the two qubits or of the two cavities. In the following we start showing how to compute for each part the time dependent density matrix elements, which in general may be different for the two subsystems.

\subsubsection{Dynamics of subsystems}
Here we consider the dynamics of a single subsystem $S$.
We choose the standard product basis ${\cal B}= \{\ket{1}=\ket{1_{\rm q}}\ket{1_{\rm c}}, \, \ket{2}=\ket{1_{\rm q}}\ket{0_{\rm c}}, \, \ket{3}=\ket{0_{\rm q}}\ket{1_{\rm c}}, \, \ket{4}=\ket{0_{\rm q}}\ket{0_{\rm c}}\}$, where $|0_{q}\rangle$ ($|0_{c}\rangle$) and $|1_{q}\rangle$ ($|1_{c}\rangle$) are the lower and upper state of the qubit (cavity).
The dynamics of qubit $q_\mathrm{S}$ under the effect of the master equation of Eq.~(\ref{master equation}) is described by the reduced density matrix
\begin{equation}\label{qubitmatrix}
  \rho^{S,q} = \left(
\begin{array}{cccc}
 \rho^{S,q}_{11}(0)P_{t} & \rho^{S,q}_{10}(0)p_{t} \\
 \rho^{S,q}_{01}(0)p^*_{t} & 1-\rho^{S,q}_{11}(0)P_{t} \\
\end{array}
\right)\, .
\end{equation}
The time dependent coefficients $P_t$ and $p_t$ can be obtained analytically, however the presence of pure dephasing gives rise to very cumbersome and lengthy equations. In this section we present analytical results only for the case $\gamma_{\rm d}= 0$.
Numerical results in presence of pure dephasing shall be included in next section. Analogously, for the cavity modes, the dynamics of the reduced density matrix can be expressed as
\begin{equation}\label{qubitmatrixc}
  \rho^{S,c} = \left(
\begin{array}{cccc}
 \rho^{S,q}_{11}(0)Q_{t} & \rho^{S,q}_{10}(0)q_{t} \\
 \rho^{S,q}_{01}(0)q^*_{t} & 1-\rho^{S,q}_{11}(0)Q_{t} \\
\end{array}
\right)\, ,
\end{equation}
where a zero-photon initial state ($\rho^{S,c}_{11}(0) = \rho^{S,c}_{10}(0)=0$ and $\rho^{S,c}_{22}(0) =1$) has been considered.
The dynamics of the reduced density matrices is obtained using the master equation expressed by Eq.~(\ref{master equation}), that is appropriate when the reservoir is at zero temperature, the coupling between the cavity and the external modes of the reservoir has a flat spectrum in the range of involved frequencies and the qubit is resonant with the cavity \cite{scala2007PRA,scala2007JPA}. We will limit our investigation to this physical condition.
In the absence of pure dephasing, from Eq.~(\ref{master equation}) one obtains,
\begin{eqnarray}\label{p and q}
p_t &=&  \mathrm{e}^{-\frac{(\gamma_{\rm c}+\gamma_{\rm q})}{4} t} \left[\,
\cos (\Omega t) + \frac{\gamma_{\rm c}-\gamma_{\rm q}}{4 \Omega} \sin(\Omega t) \,\right] , \nonumber \\
q_t &=&  \mathrm{e}^{-\frac{(\gamma_{\rm c}+\gamma_{\rm q})}{4} t} \left[\,\frac{g}
{\Omega}\, \sin(\Omega t)\,\right]\, ,
\end{eqnarray}
with  $P_t = p^2_t$ and $Q_t = q^2_t$ and where we have introduced the characteristic frequency $\Omega = \sqrt{g^2 -((\gamma_{\rm c}-\gamma_{\rm q})/4)^2}$. We point out that the function $Q_t$ is not directly connected to the decay of the cavity excited state but is linked to the exchange of the initial qubit excitation between the qubit itself, the cavity and the reservoir. In this sense, the fact that $Q_t$ goes to zero when $g=0$ reflects its dependence on the initial conditions and on the internal qubit-cavity dynamics.
Non-Markovian features in the dynamics of the subsystems occur for values of $g$ large enough to make $\Omega$ real, but such as not to compromise the validity of the RWA in our model.

We notice that these direct relationships between the functions appearing in the diagonal and non-diagonal elements of Eqs.~(\ref{qubitmatrix}) and (\ref{qubitmatrixc}) fail for $\gamma_{\rm d} \neq 0$.
Following Ref.~\cite{bellomo2007PRL}, the knowledge of two any single parts dynamics permits to obtain the dynamics of the corresponding bipartite system.

\section{Entanglement evolution}\label{sec:entanglement-evolution}

After obtaining in the previous section all the relevant dynamical coefficients, in this section we give the explicit expressions of concurrence for some couples of parties of the four-partite system. In particular we consider the two-qubit and two-cavity entanglement separately for two different initial configurations. The qubits are initially in one and two-excitation Bell-like states, while cavities are in their vacuum state.


We shall restrict our analysis of entanglement dynamics to the two-excitation entangled initial states
\begin{equation}
	\left| \Psi \right> = \left( \alpha \left| 0 0 \right >_q +
	\beta \left| 11\right >_q \right) \left| 00 \right>_c
	\equiv \left( \alpha \left| 4 \right >_q +
	\beta \left| 1 \right >_q \right) \left| 4 \right>_c\, ,
\end{equation}
where in each ket the first entry denotes a ($q$ or $c$) state of subsystem $A$, while the second entry a state of subsystem $B$.
Generalization of the results for other (eventually mixed) initial states is straightforward.

From the evolved state $\ket{\Psi_t}$ one finds the reduced density matrices of the bipartite system of interest tracing over the degrees of freedom of the noninvolved parties. We represent the density matrices in the standard computational basis $\mathcal{B}=\{\ket{1}\equiv\ket{11},\ket{2}\equiv\ket{10},\ket{3}\equiv\ket{01},\ket{4}\equiv\ket{00}\}$. In this way, the two-qubit state at time $t$ is, e.g., given by $\hat{\rho}^\Psi_{q_A q_B}(t)=$
\begin{equation}\label{two-qubitmatrix}
  \hspace{- 2 cm} \left(
\begin{array}{cccc}
  |\beta|^2 P^A_t P^B_t & 0 & 0 & \alpha\beta p^A_t p^B_t  \\
  0 & |\beta|^2 P^A_t (1-P^B_t) & 0 & 0 \\
  0 & 0 & |\beta|^2 (1-P^A_t)P^B_t & 0 \\
  \alpha\beta^\ast p^A_t p^B_t & 0 & 0 & \alpha^2+|\beta|^2 (1-P^A_t)(1-P^B_t) \\
\end{array}
\right).
\end{equation}
In the following we will consider the case of two identical subsystems, hence $p_t^A = p_t^B$.
An analogous result can be obtained for the two-cavity system density matrix $\hat{\rho}^\Psi_{c_A c_B}(t)$ just replacing $p_t$ and $P_t$ with
$q_t$ and $Q_t$ respectively.

The concurrence corresponding to the density matrix of Eq.~\ref{two-qubitmatrix} in the case of identical subsystems is found to be \cite{bellomo2007PRL}
\begin{equation} C^{q(c)}_{\Psi} =  \max \left\{ 0, 2\left| \rho^{q(c)}_{14} \right|- 2 \sqrt{\rho^{q(c)}_{22} \rho^{q(c)}_{33}}  \right\}\, .
\end{equation}

We present numerical calculations for a specific two-excitation entangled initial state $\ket{\Psi}$ with $\alpha = 0.8$.
Figure 2 displays the concurrence dynamics of the qubits $C^{q}_{\Psi}$ and of the cavity modes $C^{c}_{\Psi}$ in absence (2a) and presence (2b) of pure dephasing.
Figure 2 also displays $Q_t=\langle a^\dag a \rangle$, providing information on the detectable output photon flux $\langle a_{\rm out}^\dag a_{\rm out} \rangle = \gamma_{\rm c}\,  Q_t$.
In both cases $\gamma_{\rm c} = 0.3 g$, the other decay rates being fixed as  $\gamma_{\rm q} = 0.3 g$ and $\gamma_{\rm d} = 0$ in Fig.\ 2a while $\gamma_{\rm q} = 0$, $\gamma_{\rm d} = 0.3 g$ in Fig.\ 2b. For these values, non-Markovian features appear in each qubit-cavity subsystem dynamics, leading to similar effects in the dynamics of the plotted quantities.
The figure shows that pure dephasing affects heavily the entanglement dynamics increasing the entanglement decay and enabling  its sudden death. Although phase noise acts directly only on the emitter, owing to the strong coupling between the emitter and the cavity, it affects dramatically also the two-cavity concurrence.
\begin{figure}[!htbp]
\begin{center}
\includegraphics[height=45 mm]{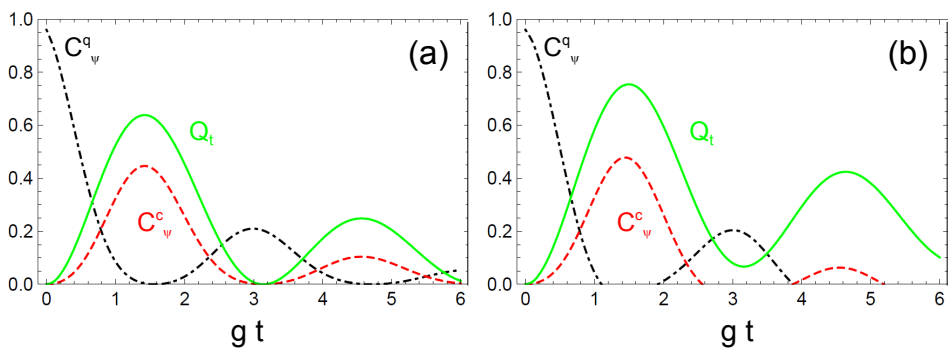}
\caption{$Q_t=\langle a^\dag a \rangle $ (green solid line), concurrences  of the qubits $C^{q}_{\Psi}$ (black dot-dashed line) and of the cavity modes $C^{c}_{\Psi}$ (red dashed line) for the initial state $\ket{\Psi }$ with $\alpha = 0.8$ and with $\gamma_{\rm c} = 0.3 g$ as a function of the dimensionless quantity $gt$ in non-Markovian regime. (a) $\gamma_{\rm q} = 0.3 g$, $\gamma_{\rm d} = 0$. (b) $\gamma_{\rm q} = 0$, $\gamma_{\rm d} = 0.3 g$. } \end{center}

\end{figure}
Figure 3a shows the two-emitter concurrence as function of time and of the amount of phase noise $\gamma_{\rm d}$.  Figure 3b displays the two-cavity concurrence $C^{c}_{\Psi}$. We used $\alpha = 1/\sqrt{2}$, $\gamma_{\rm c} = 0.17 g$, $\gamma_{\rm q} = 0$.
The detrimental effect of phase noise on both the dynamics of the entanglement of emitters and cavities is evident.
\begin{figure}[!htbp]
\begin{center}
{\includegraphics[height=50 mm]{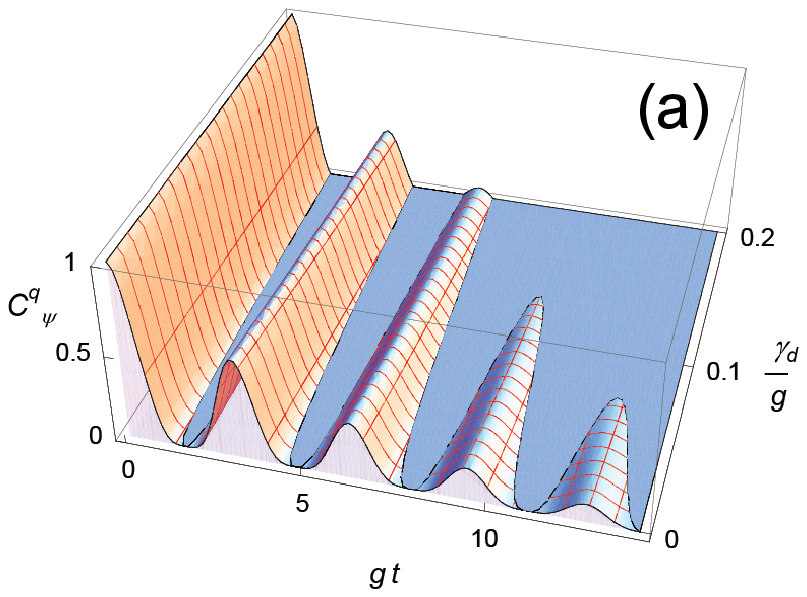}
\hspace{0.5 cm}
\includegraphics[height=50 mm]{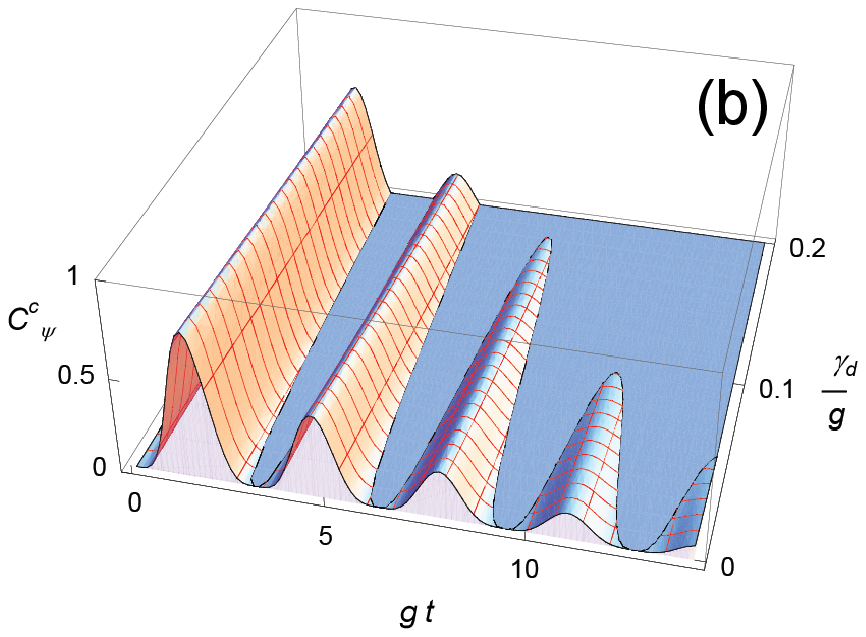}}
\caption{Concurrences of the qubits $C^{q}_{\Psi}$ (panel a) and of the two spatially separate cavity-modes $C^{c}_{\Psi}$ (panel b) as a function of the dimensionless quantities $g t$ and $\gamma_{\rm d}/g$ for the initial state $\Psi \rangle$ with $\alpha = 1/\sqrt{2}$ in non-Markovian regime for $\gamma_{\rm c} = 0.17 g$, $\gamma_{\rm q} = 0$}.
\end{center}
\end{figure}
\section{Application to quantum dots under realistic conditions}\label{par:Application}

Here we specialize to the case where the considered system is implemented by using currently available quantum dots as quantum emitters, embedded in separated micro-cavities \cite{Hennessy}.  Solid state microcavities with three-dimensional photon confinement, high Q, and small volume mode can be realized by fabricating a photonic crystal slab structure with a nanocavity composed of one or more missing air holes. The slab incorporates a central layer of low density self-assembled InAs quantum dots \cite{Reith,Hennessy}.
Another geometry of particular interest is that of  micropillar cavities \cite{Yoshie,Loo}. In these systems the fundamental cavity mode can be coupled to and from the outside with a very high coupling efficiency. Moreover, they offer interesting perspectives for the implementation of quantum information protocols using charged quantum dots \cite{Hu}.
Figure 4a displays the concurrence dynamics of the quantum emitters $C^{q}_{\Psi}$ and of the cavity modes $C^{c}_{\Psi}$ obtained for two independent cavity-embedded quantum dots ($\alpha= 0.8$). We consider typical system parameters for the state-of-art microstructures \cite{Hennessy}:
$\gamma_{\rm c}= 100\; \mu$eV, $\gamma_{\rm d}= 30 \mu$eV,$\gamma_{\rm q}= 10\; \mu$eV, $g= 110\; \mu$eV.
The figure shows that the two-dot entanglement after a rebirth survives up to about $40$ ps. Quantum operations based on all-optical implementations can be performed by means of  ultrafast pulses. At optical frequencies, pulses of 20-100 fs time-width are currently available.
Very recently structures displaying higher Q values but with a quite low coupling $g$ have been realized \cite{Loo}. Figure 4b displays the concurrence dynamics of the quantum emitters $C^{q}_{\Psi}$ and of the cavity modes $C^{c}_{\Psi}$ obtained by using $\alpha =0.8$ and parameters describing this novel micropillar structure \cite{Loo}: $\gamma_{\rm c}= 20\; \mu$eV, $\gamma_{\rm d}= 12 \mu$eV,$\gamma_{\rm q}= 4\; \mu$eV, $g= 16\; \mu$eV.
In this case both the two-dot and two-cavity modes entanglement increase their lifetime.
Figure 4 also displays the behavior of $Q_t=\langle a^\dag a \rangle$ which is proportional to the detectable output photon flux $\langle a_{\rm out}^\dag a_{\rm out} \rangle = \gamma_{\rm c}\, Q_t$.

We notice that in the systems considered in this section, dots and cavities frequencies are $\sim 1.3$ eV $\gg g$ so that the validity limits of RWA are well satisfied.

\begin{figure}[H]
\begin{center}
\includegraphics[height=50 mm]{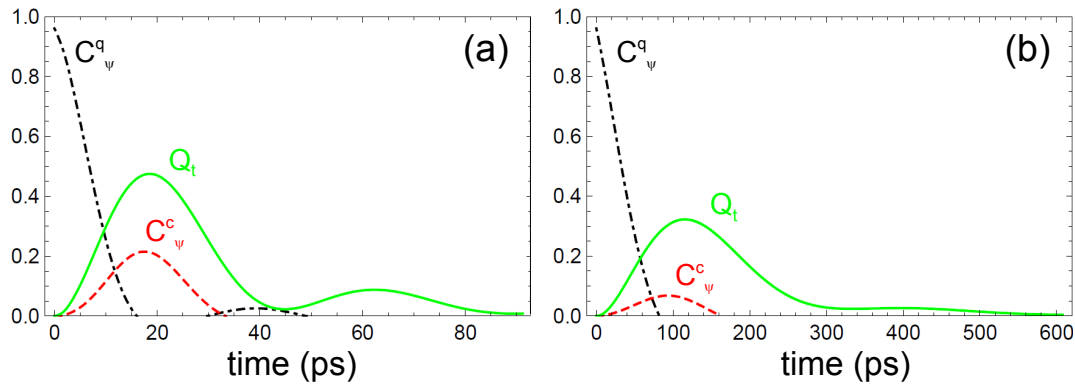}
\caption{$Q_t=\langle a^\dag a \rangle $ (green solid line), concurrences  of the quantum dots $C^{q}_{\Psi}$ (black dot-dashed line) and of the cavity modes $C^{c}_{\Psi}$ (red dashed line) for the initial state $| \Psi \rangle$ with $\alpha = 0.8$ as a function of the dimensionless quantity $gt$ in non-Markovian regime. Values of other parameters: (panel a) $\gamma_{\rm c}= 100\; \mu$eV, $\gamma_{\rm d}= 30 \mu$eV,$\gamma_{\rm q}= 10\; \mu$eV, $g= 110\; \mu$eV and (panel b) $\gamma_{\rm c}= 20\; \mu$eV, $\gamma_{\rm d}= 12 \mu$eV,$\gamma_{\rm q}= 4\; \mu$eV, $g= 16\; \mu$eV.} \end{center}
\end{figure}
 \section{Conclusions}\label{par:Conclusions}

We have studied the dynamics of entanglement in a system made by two solid-state emitters, as two quantum dots, embedded in two separated micro-cavities. In addition to the coupling with cavity mode, the emitter is subject to spontaneous emission, due to the coupling with a continuum of leaky modes, and to phonon-induced pure dephasing mechanisms. We have modeled this physical system as a multipartite system composed by two independent parts each containing a qubit exposed to cavity losses, spontaneous emission and pure dephasing.
The numerical results presented here for arbitrary values of the physical parameters put forward the impact of pure dephasing on the entanglement dynamics of the quantum emitters and of the two-cavity modes. Experimental information about the latter can be gathered by detection of the collected cavity output field. We have finally applied this theoretical framework to the specific case of currently available solid-state quantum dots in micro-cavities.

\section*{References}

\providecommand{\newblock}{}

\end{document}